\documentclass[
   10pt,pra,aps,twocolumn,showpacs,superscriptaddress]{revtex4}

\usepackage{graphicx}
\usepackage{amsmath,amssymb,revsymb}
\usepackage[colorlinks=true,citecolor=blue]{hyperref}

\begin{document}

\title{Ultracold three-body collisions near narrow Feshbach resonances}

\author{Yujun Wang}
\affiliation{Department of Physics, Kansas State University, Manhattan, Kansas, 66506, USA}
\author{J. P. D'Incao}
\affiliation{JILA, University of Colorado and NIST, Boulder, Colorado, 80309-0440, USA}
\author{B.D. Esry}
\affiliation{Department of Physics, Kansas State University, Manhattan, Kansas, 66506, USA}

\begin{abstract}
We study ultracold three-body collisions of bosons and fermions when the
interatomic interaction is tuned near a narrow Feshbach resonance. We show that the width
of the resonance has a substantial impact on the collisional properties of ultracold gases in
the strongly interacting regime.  From our numerical and analytical analyses, we identify
universal features dependent on the resonance width.
Remarkably, we find that all inelastic processes near narrow resonances leading to deeply bound
states in bosonic systems are suppressed while those for fermionic systems are enhanced.
As a result,
narrow resonances present a scenario the reverse of that found for broad resonances,
opening up the possibility of creating stable samples of ultracold bosonic gases with
large scattering lengths. 
\end{abstract}

\pacs{}
\maketitle

\section{Introduction}
Strongly interacting three-body systems play an important role in many areas of physics,
extending over condensed matter, atomic, molecular, and nuclear physics~\cite{Jensen2004,Braaten2006}. 
Advances in the control of interatomic interactions have made
ultracold atomic gases a preferred test bed for many 
interesting physical phenomena. One of the most important tools for this control is the magnetic- or optical-Feshbach 
resonance~\cite{Feshbach1962,FeshbachChin}.  Applying an external field, the 
$s$-wave scattering length $a$ between two atoms can be tuned from $-\infty$ to $+\infty$. 
Although the tunability of the interatomic interaction greatly expands the range of experimentally accessible phenomena,  a major  difficulty encountered
in the strongly interacting limit, 
$|a|$$\gg$$ r_0$ where $r_0$ is the characteristic range 
of the interatomic interactions, is that the system can become unstable
due to three-body collisional losses of atoms and molecules. 
On the other hand, a two-component Fermi gas shows extraordinary stability against few-body losses near a Feshbach resonance~\cite{DimerJochim,Jochim2003,DimerRegal2004,Regal2004,DimerPetrov2004}, which is the key ingredient that allowed the experimental realization of 
a broad range of novel physical phenomena~\cite{FermiGasGiorgini}.


The fundamental few-body physics behind the theoretical understanding of the stability of bosonic and fermionic gases is related to Efimov physics~\cite{Efimov1970,ScalingDIncao2005}. In fact, in
the past few years, few-body physics has received a great deal of attention due to the experimental verification
~\cite{Kraemer2006,4BFerlaino2009,Knoop2009,Ottenstein2008,Huckans2009,Barontini2009,Zaccanti2009,Gross2009,Pollack2009,Williams2009}
of several key features of Efimov physics, confirming our understanding of this intriguing universal phenomena. 
Nevertheless, most of the universal phenomena explored so far, except the ones in Ref.~\cite{Petrov2004,Gogolin2008,Lee2007,Smirne2007,Massignan2008,Braaten2003,Hammer2007,ReffPlatter2009}, 
have been made under the assumption that ultracold scattering properties depend only on $a$, which is expected to be a good one for broad Feshbach resonances.
This focus on broad resonances and the dominance of $a$ is natural
given the low-energy expansion of the two-body
$s$-wave scattering phase shift $\delta$,
\begin{equation}
k_2\cot\delta = -\frac{1}{a}+\frac{1}{2}r_{\rm eff} k_2^2+\ldots,
\label{Eq:PhaseShift}
\end{equation}
when $k_2\ll\sqrt{2/|a r_{\rm eff}|}$, where $k_2$ is the two-body wavenumber and $r_{\rm eff}$ is the effective range. 
For the magnetic Feshbach resonances used in ultracold experiments, $r_{\rm eff}$
can be estimated from the resonance parameters by~\cite{Petrov2004,ReffBruun2005}
\begin{eqnarray}
r_{\rm eff}=-\frac{1}{|\mu_2a_{\rm bg}\Delta\mu\Delta B|}, 
\label{reff}
\end{eqnarray}
where $\mu_2$ is the two-body reduced mass, $a_{\rm bg}$ is the background scattering length, $\Delta\mu$
is the difference of the magnetic moment between the two channels involved in the resonance, 
and $\Delta B$ is the resonance width in the magnetic field. 
The effective ranges for some selected Feshbach resonances~\cite{FeshbachChin} are listed in Table~\ref{Tab:EffRange} in App.~\ref{Append:Table}.  
Since $|r_{\rm eff}|$ is inversely proportional to the resonance width~\cite{Petrov2004}, 
the second term in Eq.~(\ref{Eq:PhaseShift}) is negligible at low collisional energies for broad resonances.
Near a narrow resonance, however, $|r_{\rm eff}|\gg r_0$ and the second term in Eq.~(\ref{Eq:PhaseShift}) 
is no longer negligible for three-body collisions even at ultracold collision energies.
This implies that $r_{\rm eff}$ should 
be incorporated in the three-body universal theory for the narrow resonances. 

Although some authors regard three-body ``universality'' as the determination of three-body properties by the scattering length alone, 
universality can be more broadly defined to include the dependence of three-body properties on 
a few parameters that encapsulate the details of the short-range interactions. 
As we will show here, the role of a large effective range is similar to that of the scattering length in the sense that it {\em universally} affects the behavior of the  three-body system. 

Narrow resonances are expected not only to affect the behavior underlying Efimov physics~\cite{Petrov2004,Gogolin2008} 
but also have consequences for the BEC-BCS crossover picture for fermionic systems~\cite{DePalo2004}.  In fact, the possibility
of new many-body physics near narrow Feshbach resonances has been
proposed theoretically in Refs.~\cite{Sorensen2009,Sorensen20092}.
Furthermore, the closed-channel-dominant nature of a narrow Feshbach resonances may give novel few-body physics, since
the two-body wavefunction always carry strong bound state
characteristics even if they do not form a bound state.
Therefore, understanding the role of the effective range is crucial. 
While specific systems have been modeled near a resonance~\cite{Kartavtsev2002,Smirne2007,Massignan2008,Mehta2008},
no simple analytical expressions for scattering observables near narrow resonances, like those for broad resonances~\cite{Nielsen1999,Esry1999,Braaten2006}, have yet  been obtained.

We believe that a better understanding of the physics near narrow resonances becomes increasingly important
since narrow Feshbach resonances are frequently encountered in gases with  mixed atomic
species in many recent experiments ~\cite{Stan2004,Wille2008,Voigt2009,Pilch2009}. 
Since the scattering lengths and effective ranges in a heteronuclear system can take quite different values for distinct pairs by tuning through overlapping resonances,
a rich variety of three-body physics is expected to arise in such systems.
And, the development
of optical Feshbach resonances techniques~\cite{OptResBohn1997,OptResFatemi2000,OptResTheis2004} promises
greater experimental control over the width of the resonance.
It is thus important to develop the universal theory for the three-body physics near narrow resonances.

Our goal with this paper is to provide the general framework for such a theory and derive explicit results for several cases. 
We are especially interested in the change of the universal behavior for the three-body processes
that depend on the short-range physics. 
In particular, we study the dependence of ultracold three-body collision rates on $r_{\rm eff}$,
obtaining analytic expressions for $K_3$ and $V_{\rm rel}$ similar
to those already obtained for broad resonances [see Eqs.~(\ref{Eq:K3:aN}) and (\ref{Eq:Vrel:aP})]. 
Moreover, we verify these expressions with numerical solutions. and verified by numerical solutions.
Surprisingly, we find that {\em all} inelastic processes leading to
deeply-bound two-boson states are suppressed as $|r_{\rm eff}|^{-1}$, indicating that bosonic gases
are more stable near narrow resonances. In contrast, we find that fermionic gases
should be less stable due to enhanced losses for large $|r_{\rm eff}|$. 
We use atomic units (a.u.)
throughout this paper unless otherwise stated.

\section{Theory}

\subsection{Background}
\label{background}


In the previous work on this topic by Petrov~\cite{Petrov2004} and Gogolin, {\em et al}~\cite{Gogolin2008}, 
the zero-range potential (ZRP)~\cite{ZRP} model was used, making it difficult to evaluate the importance of
the short-range physics.  In particular, Petrov calculated the 
recombination rate for identical bosons near a narrow Feshbach resonance 
with $a$$>$0 using a modified ZRP to include $r_{\rm eff}$~\cite{Petrov2004}.
By solving his model numerically, he found that the minima described by Eq.~(\ref{Eq:K3:aP}) still appear
for $a$$\gg$$|r_{\rm eff}|$ but occur at fixed values of $|r_{\rm eff}|$/$a$ without any reference to the three-body
parameter $\Phi$.  Gogolin {\it et al.}~\cite{Gogolin2008} reproduced these results with a very 
different method but effectively the same physical model.
Other theoretical efforts seek to address the modification of Efimov physics by the effective range~\cite{Braaten2003,Hammer2007,ReffPlatter2009} but are restricted to
$|r_{\rm eff}|\approx r_0$ such that their perturbative treatment remains adequate. Their results, therefore, cannot be applied to
narrow Feshbach resonances.

More fundamentally, and in contrast to \cite{Petrov2004} and \cite{Gogolin2008}, we show that 
the short-range three-body physics {\em is} important near a narrow resonance for identical bosons
--- even in the limit $r_0$$\rightarrow$0.  
Petrov and Gogolin {\it et al.} took advantage of the fact that the inclusion of $r_{\rm eff}$ in the ZRP model regularizes the three-body dynamics, eliminating the 
mathematical need for a three-body short-range parameter. This regularization had been investigated earlier in Ref~\cite{Fedorov2001} and was recently revisited in \cite{Thogersen2009}.
While the Thomas collapse~\cite{Thomas1935} is avoided, we will show that 
a three-body parameter is still needed to represent the short-range three-body physics and that, consequently, $a$ and $r_{\rm eff}$ alone are insufficient
to describe ultracold three-body observables.

\subsection{Modeling Feshbach resonances}

\begin{figure}[h]
\begin{center}
\includegraphics[height=1.25in]{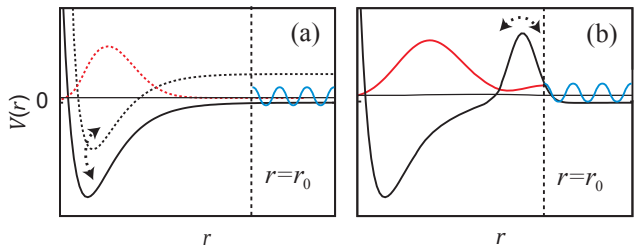} 
\end{center}
\caption[Comparison between a Feshbach resonance and a shape resonance.]
{Schematic comparison of a Feshbach resonance (a) and a shape resonance (b). On resonance, the short-range ($r<r_0$), bound component is similar in both cases and the long-range ($r>r_0$) 
wavefunction is the same. The double-headed arrows indicate the coupling.}
\label{Fig:FeshbachModels}
\end{figure}

Feshbach resonances~\cite{FeshbachChin} are a multi-channel phenomenon.
In general, a two-channel model is sufficient to reproduce the main resonant structure contained in the real problem.
Roughly speaking, a Feshbach resonance occurs when a state of  the closed (upper) channel becomes degenerate with the collision energy of two free atoms in the open (lower) channel, 
as illustrated in Fig.~\ref{Fig:FeshbachModels}(a). 
In ultracold gases, the tunability afforded by a magnetic Feshbach resonance of interactions is achieved by tuning the separation of 
the asymptotic thresholds of the two channels such that the scattering length in the lower channel goes through a pole
when the closed channel state moves across the threshold of the open channel. The width of the resonance, and therefore the effective range [see Eq.~(\ref{reff})], 
are mainly controlled by the strength of the coupling between the open and closed channels. 
Generally, strong coupling produces broad Feshbach resonances (small $|r_{\rm eff}|$);
and weak coupling, narrow resonances (large $|r_{\rm eff}|$).

If there is only one open channel, one can design a single channel model that has the same asymptotic scattering wavefunction as the multi-channel wavefunction at 
ultracold collision energies. Such a model is possible since the asymptotic wavefunction does not depend on 
the short-range two-body physics that generates the resonance~\cite{Petrov2004,Jonsell2004}. 
We thus model the Feshbach resonance with a single-channel potential that supports a shape resonance as shown in Fig.~\ref{Fig:FeshbachModels}(b).
Specifically, we use
\begin{eqnarray}
V_{\rm sech}(r)=-D\mathrm{sech}^2(3r/r_0)+B e^{-2(3r/r_0-2)^2},
\label{Eq:2BPotential}
\end{eqnarray}
where $r$ is the distance between the two atoms. The potential depth $D$ primarily controls $a$, and the barrier height $B$ is adjusted to produce the desired $r_{\rm eff}$.  
Instead of coupling two different channels as in the Feshbach resonance case, our model 
couples the scattering region ($r>r_{0}$) with the short-range region ($r<r_{0}$).

In order to investigate the three-body universality, i.e., universal properties that do not depend on a particular choice of model interaction, 
we have also used a two-body potential consisting of the sum of a Morse potential and a Gaussian barrier, given by
\begin{align}
V_{\rm Morse}(r)\!=\!D[(1\!-\!e^{-(3r/r_0-1)})^2\!-\!1]\!+\!B e^{-2(3r/r_0-2)^2}.
\label{Eq:2BMorse}
\end{align}

\subsection{Adiabatic hyperspherical representation}

We solve the three-body Schr{\"o}dinger equation using the adiabatic hyperspherical representation~\cite{Suno2003}.
In this representation, the overall size of the three-body system is characterized by the hyperradius $R$, and the three-body 
configuration is represented by a set of hyperangles $\Omega$.

After scaling the three-body wavefunction $\Psi$ by $\psi=R^{5/2}\Psi$, the Schr{\"o}dinger equation can be written as
\begin{eqnarray}
\left[-\frac{1}{2\mu}\frac{\partial^2}{\partial R^2}+H_{\mathrm{ad}}(R;\theta,\varphi)\right]\psi=E\psi,
\label{Eq_Sch}
\end{eqnarray}
where $E$ is the total energy, $\mu=m/\sqrt{3}$ is the three-body reduced mass.
We solve Eq.~(\ref{Eq_Sch}) by first expanding the three-body wavefunction as
\begin{eqnarray}
\psi=\sum_{\nu=0}^{\infty} F_{\nu E}(R)\Phi_\nu(R;\Omega),
\end{eqnarray}
where the adiabatic channel functions $\Phi_\nu(R;\Omega)$ are solutions of the adiabatic equation
\begin{eqnarray}
H_{\rm ad}\Phi_\nu=U_\nu(R)\Phi_\nu,
\label{Eq_Adiabatic}
\end{eqnarray}
solved for fixed values of $R$. Here $U_{\nu}(R)$ is the adiabatic hyperspherical potential.
Upon substitution of $\psi$, Eq.~(\ref{Eq_Sch}) reduces to a set of coupled one-dimensional equations:
\begin{eqnarray}
&&\!\!\!\!\!\!\!\!\left[-\frac{1}{2\mu}\frac{d^2}{d R^2}+U_{\nu}(R)\right]\!F_{\nu,E}(R)
\!-\!\frac{1}{2\mu_3}\sum_{\nu'}\!\left[2P_{\nu,\nu'}(R)\frac{d}{dR}\right.\nonumber\\
&&\!\!\!+Q_{\nu,\nu'}(R)\bigg]F_{\nu',E}(R)\!=\!E F_{\nu,E}(R),
\label{Eq_Radial}
\end{eqnarray}
with non-adiabatic couplings $P_{\nu,\nu'}$ and $Q_{\nu,\nu'}$ defined by
\begin{eqnarray}
P_{\nu,\nu'}(R)=\left\langle\!\!\!\left\langle\Phi_{\nu}\left|\frac{d}{d R}\right|\Phi_{\nu'}\right\rangle\!\!\!\right\rangle,\\
Q_{\nu,\nu'}(R)=\left\langle\!\!\!\left\langle\Phi_{\nu}\left|\frac{d^2}{d R^2}\right|\Phi_{\nu'}\right\rangle\!\!\!\right\rangle.
\end{eqnarray}
Here, the double brackets denote integration over only the hyperangular degrees of freedom.
The hyperspherical adiabatic representation, therefore, offers a simple and conceptually clear description of three-body scattering processes. The non-adiabatic couplings drive the inelastic collisions between channels characterized by the effective potentials 
\begin{eqnarray}
W_{\nu,\nu}(R)=U_\nu(R)-\frac{1}{2\mu}Q_{\nu,\nu}(R),
\label{Eq:EffPot}
\end{eqnarray}
which in turn support all bound states and resonances of the system and dictate many of the scattering properties of the system.
Here, we obtain the scattering observables by solving Eq.~(\ref{Eq_Radial}) using the $R$-matrix method~\cite{Aymar1996}, 
while the analytical properties are derived primarily based on the properties of the effective potentials.

\section{Results and discussion}
\label{Section:Narrow:Bosons}

\subsection{Three-body inelastic processes for identical bosons}

For an atomic sample with spin-stretched bosons $B$ in their lowest hyperfine state, 
it is well known that the atomic losses are dominated by three-body recombination, $B$+$B$+$B$$\rightarrow$$B_2$+$B$, 
releasing enough kinetic energy for collision products to escape from the trap. 
For $|a|$$\gg$$r_0$ near a broad Feshbach resonance,
the three-body recombination rates $K_3$ have the following universal expressions~\cite{Nielsen1999,Esry1999,Braaten2006}:
\begin{align}
\label{Eq:K3:aP}
K_3^{(a>0)}&\!=\! 67.1 e^{-2\eta}\!\left(\sin^2[s_0\ln\frac{a}{r_0}\!+\!\Phi]\!+\!\sinh^2\eta\right)\!\frac{a^4}{m},\\
\label{Eq:K3:aN}
K_3^{(a<0)}&\!=\!\frac{4590\sinh 2\eta}{\sin^2[s_0\ln(\lvert a\rvert/r_0)\!+\!\Phi\!+\!1.53]\!+\!\sinh^2\eta} \frac{a^4}{m},
\end{align}
where $s_0$$\approx$1.00624, $\Phi$ is a short-range three-body phase, and $\eta$ is a parameter that represents the probability 
of inelastic transitions to deeply bound dimers at short distances. 
While the overall behavior of $K_3$ is determined by $a$, any comparison with experimental data requires the short-range three-body parameters
 $\Phi$ and $\eta$ to be properly determined.
In general, these three-body parameters can not be predicted from two-body physics alone~\cite{PhaseDIncao}, and must be determined from some three-body observable, 
usually by fitting.

If weakly-bound molecules $B_2^*$ are present, their lifetime is determined by 
atom-dimer relaxation processes, 
$B_2^*$+$B$$\rightarrow$$B_2$+$B$~\cite{Braaten2006}. The vibrational relaxation rate
$V_{\rm rel}$ also has a universal form when $a>0$ and $a\gg r_0$:
\begin{equation}
\label{Eq:Vrel:aP}
V_{\rm rel}=\frac{20.3 \sinh 2\eta}{\sin^2[s_0\ln(a/r_0)\!+\!\Phi\!+\!1.47]\!+\!\sinh^2\eta}\frac{a}{m}.
\end{equation}
When $a<0$, $V_{\rm rel}$ is independent of $a$.

\subsubsection{Numerical inelastic rates}
One implicit, reasonable, assumption behind the universal expressions Eqs.~(\ref{Eq:K3:aP}) and (\ref{Eq:K3:aN}) 
is that the three-body potentials and couplings at short distances ($R\lesssim r_0$) remain practically unchanged when the scattering length is tuned through a pole. 
Similarly, to extract the universal behavior of a three-body system with large effective range, 
we also want to have negligible changes in the short-range physics when the effective range is tuned.
Otherwise, the universal scaling behavior will be unphysically entangled with the non-universal changes of the short-range physics.
With $V_{\rm sech}$ and $V_{\rm Morse}$, however, a non-negligible change in the barrier height is required to make $a$ large when $r_{\rm eff}$ is changed through a large range. 
This leads to non-negligible changes for the short-range physics.
To solve this problem, we 
introduced a hard wall in $W_{\nu,\nu}(R)$ at $R$=$r_0$ to cut-off the non-universal short-range behavior. The behavior of the potentials
$W_{\nu,\nu}(R)$ beyond $R$=$r_0$ are universal, allowing us to extract the universal properties of the scattering observables.
To confirm that the insertion of the hard wall at $R=r_{0}$ does not lead to undesired effects, we have verified that our numerical results are not sensitive to the precise location of this hard wall.

\begin{figure}
\begin{center}
\includegraphics[height=3.2in]{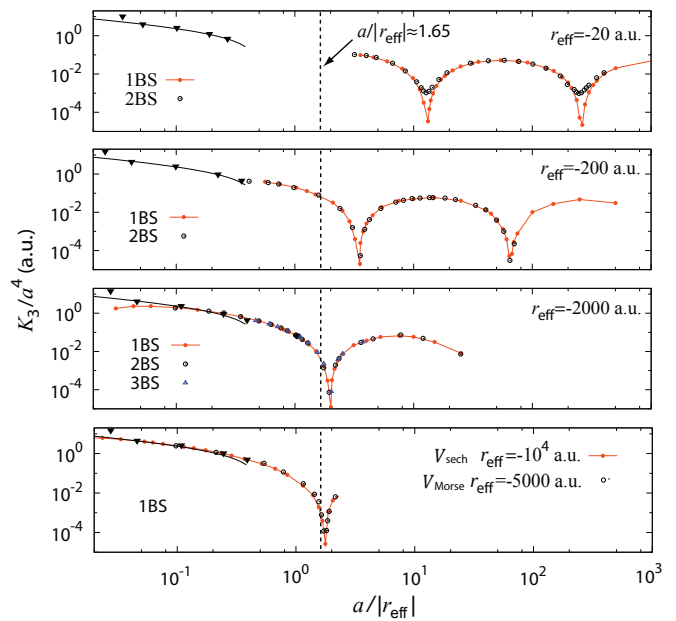} 
\end{center}
\caption[Ultracold three-body recombination rates ($a>0$) for identical bosons.]
{\small{Ultracold three-body recombination rates for identical bosons ($a>0$). 
The vertical dashed lines indicate the position of the first Efimov feature predicted in Refs.~\cite{Petrov2004,Gogolin2008}.
The solid triangles are experimental data for the 907G Feshbach resonance in a Na BEC~\cite{Stenger1999}.
Here we show $K_3$ with one or more two-body $s$-wave bound states (BS),
including the analytical result from \cite{Petrov2004} scaled to match our data (thick solid line). In the lowest panel we
show the recombination rates calculated by using two different two-body potentials: the potential defined in Eq.~(\ref{Eq:2BPotential})
and (\ref{Eq:2BMorse}).
Here $r_0$=50~a.u..}
}
\label{Fig:K3:aP:Bosons}
\end{figure}

Figure~\ref{Fig:K3:aP:Bosons} shows our numerical calculations of $K_{3}$ when $a>0$.
We generated each curve by varying $a$ at fixed $|r_{\mathrm{eff}}|$, mimicing 
the tuning of magnetic field across a Feshbach resonance for the cases where $|r_{\mathrm{eff}}|$ does not change significantly across the resonance. 
For $a$$\gg$$|r_{\rm eff}|$, the rates retain
the features predicted in Eq.~(\ref{Eq:K3:aP}),  due to the fact that the scaling behavior of the inelastic rates
should still be determined by Efimov physics when $a$ is the largest length scale in the system. 
For $a$$\lesssim$$|r_{\rm eff}|$, however, $K_3$ deviates from this formula and approaches the
$(a^7|r_{\mathrm{eff}}|)^{1/2}$ behavior predicted in \cite{Petrov2004} [see black solid line in Fig.~\ref{Fig:K3:aP:Bosons}].  The figure also shows that
the rates seem to converge to a universal curve as $|r_{\rm eff}|$ increases. 
Moreover, as the limit $r_0/|r_{\rm eff}|\rightarrow 0$ is approached, the position of the first Efimov minimum as a function of $|a|$/$|r_{\rm eff}|$ agrees
reasonably well with the ZRP 
predictions from Refs.~\cite{Petrov2004} and \cite{Gogolin2008}.
 However, as will be discussed later, this agreement should not be expected in general.
In any case, Table~\ref{Tab:EffRange} shows that $r_{\rm eff}$ is typically below $10^3$~a.u. for
those resonances with a few gauss of width. Therefore typical experiments will actually fall far from range of validity of the ZRP results since $r_0\sim 100$a.u.. 
Thus, for the moderate values of $|r_{\rm eff}|$ in typical experiments, Fig.~\ref{Fig:K3:aP:Bosons} shows that the positions of the minima in $a/|r_{\rm eff}|$ deviate dramatically from the ZRP prediction. 
It followed that any experimental attempt to reduce atomic
losses by tuning the ratio $a/|r_{\rm eff}|$ close to a minimum in $K_3$ will require knowledge of the short-range three-body physics just as for broad resonances. 

One interesting observation in Fig.~\ref{Fig:K3:aP:Bosons} is that the minima in $K_3$ for more than one two-body bound state become
more pronounced, i.e., the value of $K_{3}$ at a minimum decreases as $|r_{\rm eff}|$ increases. If there were a single weakly bound two-body state, 
$K_3$ would vanish at the minimum.
When deeply bound states are available to recombine into, however, $K_3$ no longer vanishes at the minima.
This situation can be clearly seen in Fig.~\ref{Fig:K3:aP:Bosons} when  $|r_{\rm eff}|\approx r_{0}$. 
The deeper minima for larger $|r_{\rm eff}|$, therefore, indicate the suppression of recombination into deeply-bound states.
As will be discussed later, this suppression is the consequence of a new scaling behavior 
with large $|r_{\rm eff}|$.

\begin{figure}
\begin{center}
\includegraphics[height=3.2in]{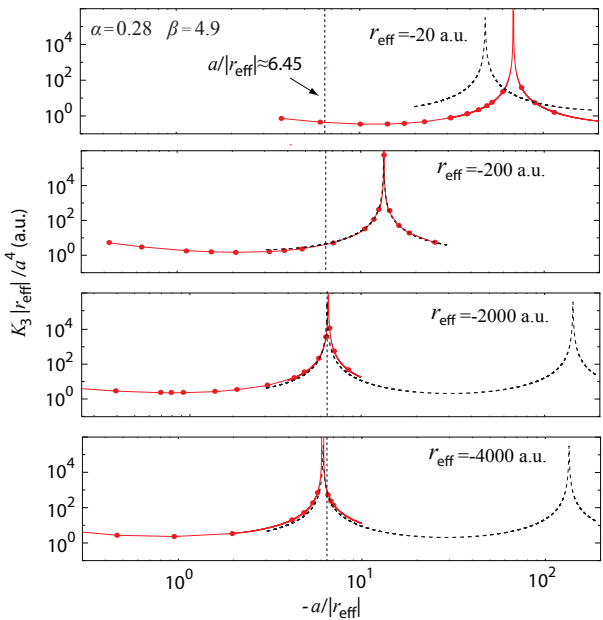} 
\end{center}
\caption[Ultracold three-body recombination rates ($a<0$) for identical bosons. ]
{\small{Ultracold three-body recombination rates ($a<0$) for identical bosons. 
The solid lines through the numerical data points are to guide the eye, while the dashed lines are the analytical results
from Eq.~(\ref{Eq:K3:aN:Narrow}), using the $\alpha$ and $\beta$ indicated 
from a fit in the limit $|r_{\rm eff}|$$\rightarrow$$\infty$.  
Here $r_0$=50~a.u..}
}
\label{Fig:K3:aN:Bosons}
\end{figure}

In Fig.~\ref{Fig:K3:aN:Bosons}, we show $K_3$ for $a<0$. The rates are scaled as $K_3|r_{\rm eff}|/a^4$ to show the overall scaling with $|r_{\rm eff}|$.
Similar to $K_3$ for $a>0$, the numerical rates are calculated by changing $a$ with a fixed $r_{\rm eff}$. When $|r_{\rm eff}|$
increases, the curves  converge to a limiting case, and we compare the positions for the first peak in $K_3$ with its ZRP prediction.
In particular, the $K_3$ prediction is based on the scattering length when the first Efimov state becomes bound~\cite{Gogolin2008}.
It can be seen that this prediction is close to our limiting position, but is not in exact agreement.

\begin{figure}
\begin{center}
\includegraphics[height=2.7in]{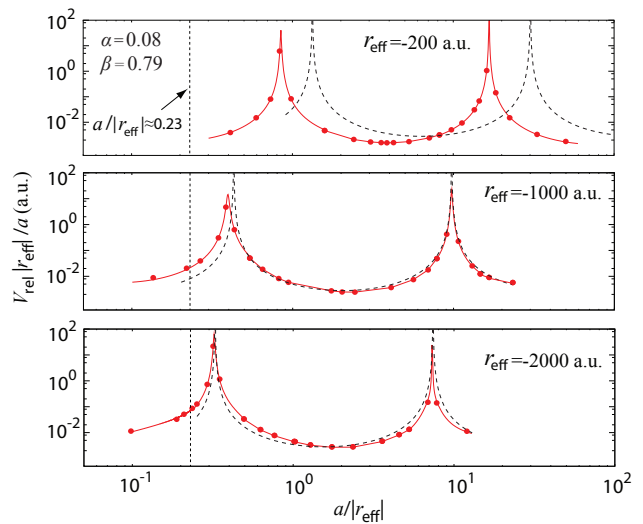} 
\end{center}
\caption[Ultracold three-body relaxation rates ($a>0$) for identical bosons. ]
{\small{Ultracold three-body relaxation rates ($a>0$) for identical bosons. 
The solid lines through the numerical data points are to guide the eye, while the dashed lines are the analytical results
from Eqs.~(\ref{Eq:Vrel:aP:Narrow}), respectively, using the $\alpha$ and $\beta$ indicated 
from a fit in the limit $|r_{\rm eff}|$$\rightarrow$$\infty$.  
Here $r_0$=50~a.u..}
}
\label{Fig:Vrel:Bosons}
\end{figure}

In Fig.~\ref{Fig:Vrel:Bosons}, we show the scaled numerical relaxation rate $V_{\rm rel}|r_{\rm eff}|/a$ for $a>0$, calculated by changing $a$ with a fixed $r_{\rm eff}$.
The curves again show convergent behavior to a limiting case, 
and the prediction on the position for the first peak is
based on the position of the first pole in the atom-dimer scattering length in \cite{Petrov2004}. The position in our numerical calculation is seemingly approaching 
the prediction, but the agreement is not clear.
One important feature in Figs.~\ref{Fig:K3:aN:Bosons} and \ref{Fig:Vrel:Bosons} is that the rates for each inelastic process have similar 
magnitude when multiplied by $|r_{\rm eff}|$, which indicates a $1/|r_{\rm eff}|$ suppression in both of the inelastic processes.

\subsubsection{Adiabatic potentials and analytical results}
\label{Potentials}
A common property
of there-body recombination ($a<0$) and relaxation ($a>0$) is that the couplings between the initial and final channels are significant only at short range. 
Inelastic transitions for these processes thus predominantly occur when $R\lesssim r_0$. 
To understand the $1/|r_{\rm eff}|$ scaling behavior, we study the properties of the corresponding adiabatic hyperspherical potentials.
This approach has proven successful for processes near broad resonances since the adiabatic potentials and their couplings have proven to be universal~\cite{LongDIncao2005}.
In App.~\ref{Chap:Narrow:ZRP}, we discuss the scaling behavior of the adiabatic potentials in ZRP model, 
and show the inadequacy of this model in deriving the adiabatic potentials for narrow resonances. In this section, we will discuss scaling behavior of the potentials based on our numerical calculations.

\begin{figure}
\begin{center}
\includegraphics[height=2.4in]{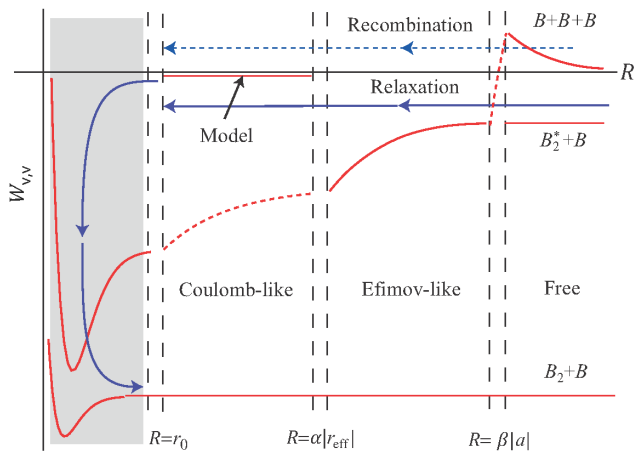}
\end{center}
\caption[Schematic adiabatic hyperspherical potentials near a narrow Feshbach resonance.]{
{Schematic
$W_{\nu,\nu}(R)$ used to derive Eq.~(\ref{Eq:Vrel:aP:Narrow}) and (\ref{Eq:K3:aN:Narrow})
with $r_0$$\ll$$ |r_{\mathrm{eff}}|$$\ll$$ |a|$. The difference in the potentials between the recombination ($a<0$) and relaxation ($a>0$) is only in the asymptotic region.
In the ``Coulomb-like'' region, the potential shown in dashed line is attractive only when there is one or two two-body bound states in the system.
The short-range region is shadowed 
to indicate strong coupling between the initial and final channels, where inelastic transition occur dominantly. }}
\label{Fig:NarrowPots:Ideal}
\end{figure}

\begin{figure}[!h]
\begin{center}
\includegraphics[height=2.2in]{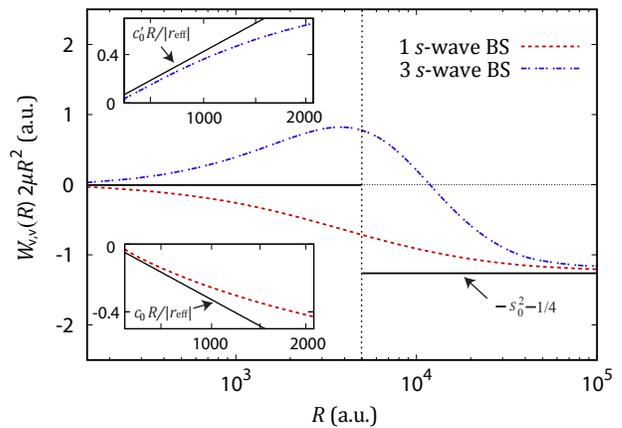}
\end{center}
\caption[Adiabatic hyperspherical potentials near a narrow Feshbach resonance.]{
\small{Adiabatic hyperspherical potentials near a narrow Feshbach resonance.
Numerical $W_{\nu,\nu}(R)$ multiplied
by 2$\mu R^2$ with $r_{\rm eff}$=--5000~a.u. and $|a|$=$\infty$. The lower curve in from the calculations
with a single $2+1$ $s$-wave bound channel and the upper one is from the calculations
with three $2+1$ $s$-wave bound channel.
}}
\label{Fig:NarrowPots}
\end{figure}

In the following we study the universal scaling of the adiabatic potentials and their couplings for $|a|\gg|r_{\rm eff}|$. The scalings for $|a|\ll|r_{\rm eff}|$ is more involved and will not be 
discussed in this paper.
Figure~\ref{Fig:NarrowPots:Ideal} shows the idealized $W_{\nu,\nu}(R)$ for three identical bosons with
$r_0\ll |r_{\rm eff}|\ll |a| $. The potentials are exactly the same as for a broad resonance except in the range $r_0\ll R\ll |r_{\rm eff}|$. In this range, $W_{\nu,\nu}(R)$ for the weakly-bound 
takes the Coulomb-like form~\cite{Petrov2004,Petrov2005}:
\begin{eqnarray}
W_{\nu,\nu}=\frac{c_0}{2\mu_3 |r_{\mathrm{eff}}|R}
\label{ReffPot}
\end{eqnarray}
instead of the usual attractive $1/R^2$ Efimov potential.  
The Coulomb-like behavior is what remains after a cancellation of the $1/R^2$ leading order terms in the potential $U_\nu(R)$ by the diagonal correction
$-Q_{\nu,\nu}/2\mu$ in Eq.~(\ref{Eq:EffPot}). 
For broad Feshbach resonances, the diagonal correction $Q_{\nu,\nu}/2\mu$ is  proportional to $1/R^3$
in the region $r_0\ll R\ll |a|$~\cite{LongDIncao2005} and as a result it does not cancel the attractive $1/R^2$ coming from $U_\nu(R)$. 
For narrow resonances, although the adiabatic potentials $U_\nu(R)$ still have the Efimov behavior when $r_0\ll R\ll |a|$,
$Q_{\nu,\nu}/2\mu$ is instead proportional to $1/R^2$ in the region $r_{0}$$\ll$$ R$$ \ll$$ |r_{\rm eff}|$ and surprisingly washes away the Efimov behavior in
the potential in this region. Studies~\cite{Thogersen2009} have shown that this new behavior
in $Q_{\nu,\nu}/2\mu$ comes from a non-universal short-range component of the two-body wavefunction in the regions where two atoms are close to each other, which is absent in 
the ZRP treatments. 
Indeed, our numerical analysis indicates that the coefficient $c_0$ in Eq.~(\ref{ReffPot})
is not universal --- it can change from positive to negative values depending on the number of deeply bound states (see Fig.~\ref{Fig:NarrowPots}).  
Figure~\ref{Fig:NarrowPots} shows that the potentials in the region $R$$\ll$$|a|$ are not universal. 
When $R$$\gg$$|r_{\rm eff}|$ the potentials recover the universal Efimov behavior.
This non-universality of the three-body effective potentials makes it
surprising that the inelastic rates in Figs.~\ref{Fig:K3:aP:Bosons}--\ref{Fig:Vrel:Bosons}, calculated with different model potentials and different number of two-body bound states, are universal.

To better understand the origin of this universal behavior, we use our idealized potentials $W_{\nu,\nu}(R)$ shown in Fig.~\ref{Fig:NarrowPots:Ideal} 
to derive analytic expressions for the collision rates. 
Based on our experience on analyzing broad resonances, the inelastic rates from full numerical calculations are expected to agree with the analytical expressions
when the prerequisites of the universal scaling hold: $r_0\ll |r_{\rm eff}|\ll |a|$. Empirically, the relation ``$\ll$'' here typically means more than one order of magnitude,
which is similar to the prerequisite $r_0\ll |a|$ for broad Feshbach resonances~\cite{LongDIncao2005}.
For the purpose of deriving simple analytic expressions, we use $\alpha|r_{\rm eff}|$ and $\beta |a|$~\cite{Jonsell2006,Mehta2009} to quantify the boundaries of the hyperradial regions, 
as shown in Fig.~\ref{Fig:NarrowPots:Ideal}.
The parameters $\alpha$ and $\beta$ will be determined by fitting our final expression to the numerical results. 
As shown in Fig.~\ref{Fig:NarrowPots:Ideal}, the non-universal region $r_0\ll R\ll |r_{\rm eff}|$ requires special handling. In principle, we can retain the Coulomb-like potential.
But, recognizing that its effective charge $c_0/2\mu|r_{\rm eff}|$ is small and  
the sign/value of $c_0$  does not affect the numerical rates, 
we simply set the potential in our model to be zero in the region $r_0\ll R\ll |r_{\rm eff}|$.
Therefore, the only effect of the existence of the Coulomb-like region is to physically separate the short-range region and the Efimov region.

As with any analytic treatment of universal three-body processes, the short-range three-body physics must be parametrized. For instance,
in Eqs.~(\ref{Eq:K3:aP})--(\ref{Eq:Vrel:aP}) for broad resonances, the three-body physics for $R\lesssim r_0$ is parametrized by $\Phi$ and $\eta$. 
For the narrow resonance considered here, we will instead 
use a complex three-body short-range scattering length $A$, 
where the real part ${\rm Re}A$ characterizes
the low-energy elastic scattering  and the imaginary part ${\rm Im}A$ accounts for the short-range inelastic transitions to deeply-bound two-body states. 
This choice permits us to write the final expressions in a form that most closely resembles Eqs.~(\ref{Eq:K3:aN})--(\ref{Eq:Vrel:aP}).
We calculate $V_{\rm rel}$ by considering 
incidence in the weakly-bound atom-dimer channel
with the transition to the deeper two-body channels driven by the non-adiabatic coupling 
localized in the region $R$$\lesssim$$r_0$.  
The derivation of the analytic rates is detailed in App.~\ref{Append:Analytic}. Briefly, the hyperradial wavefunction is written down in each region, 
then matched at the boundaries.
The relaxation rate for identical bosons is then obtained from $V_{\rm rel}^{(B)}$=$\pi (1-{\cal R})/\mu k$ where
$\cal R$ is the elastic scattering probability, yielding
\begin{equation}
V_{\rm rel}^{(B)}=
\frac{2\sqrt{3}\pi\beta \sin2\varphi_0\sinh2\eta}
{\sin^2[s_0\ln(|a/r_{\mathrm{eff}}|)+\Phi+\varphi]+\sinh^2\eta}
\frac{a}{m}
\label{Eq:Vrel:aP:Narrow}
\end{equation}
where 
\begin{align}
\tan\Phi&=2s_0\frac{\alpha-{\rm Re}A/|r_{\mathrm{eff}}|}{\alpha+{\rm Re}A/|r_{\mathrm{eff}}|},\\
\sinh\eta&=\left|\frac{{\rm Im} A}{\alpha r_{\mathrm{eff}}}\right|\csc(2\varphi_0)\sin^2(\Phi+\varphi_0), 
\label{Eq:Narrow:Phi}
\end{align}
and 
\begin{eqnarray}
\varphi&=&s_0\ln(\beta/\alpha)+\varphi_0,\\
\tan\varphi_0&=& \frac{s_0}{l+1/2}.
\label{Eq:Narrow:Varphi} 
\end{eqnarray}
The effective angular momentum~\cite{Esry2001} refers to the initial channel, and is $l=0$ for relaxation. 

With a virtually identical analysis, $K_3^{(a<0)}$ for $r_0\ll \alpha |r_{\rm eff}|\ll \beta |a|$ can be derived to be 
\begin{equation}
K_3^{(a<0)}=
\frac{12\sqrt{3}\pi^3\beta^4 \sin2\varphi_0\sinh2\eta}
{\sin^2[s_0\ln(|a/r_{\mathrm{eff}}|)+\Phi+\varphi]+\sinh^2\eta}
\frac{a^4}{m}.
\label{Eq:K3:aN:Narrow}
\end{equation}
Note that since different Hamiltonian apply --- one for $a<0$ and one for $a>0$ --- $\alpha$ and $\beta$ can take different values for $V_{\rm rel}^{(B)}$ and $K_{3}$. 
The effective angular momentum in this case is $l=3/2$~\cite{Esry2001} for the lowest three-body continuum channel.  Similar expressions can be derived for other low-energy
scattering observables. 

In Figs.~\ref{Fig:K3:aN:Bosons} and \ref{Fig:Vrel:Bosons}, the values of $\alpha$ and $\beta$ are fitted from the $|r_{\rm eff}|$-dependence of the peak positions in 
the numerical $V_{\rm rel}$ and $K_3^{(a<0)}$, in the universal limit $|r_{\rm eff}|\gg r_0$. The analytic results show converging behavior to the numerical calculations in the universal limit
$|r_{\rm eff}|\gg r_0$. One interesting point is that the universal limit is approached at smaller value of $|r_{\rm eff}|\gg r_0$ for $K_3^{(a<0)}$ than for $V_{\rm rel}$.
For $K_3^{(a<0)}$, the analytic results is almost on top of the numerical ones when $|r_{\rm eff}|/r_0$=$4$, whereas significant deviation is observed in $V_{\rm rel}$
at the same value of $|r_{\rm eff}|/r_0$.

The comparison of Eqs~(\ref{Eq:Vrel:aP:Narrow}) and (\ref{Eq:K3:aN:Narrow}) with Eqs.~(\ref{Eq:Vrel:aP}) and (\ref{Eq:K3:aN}),
respectively, shows that $r_0$ is replaced by $\alpha|r_{\rm eff}|$, as one would expect with the introduction 
of a new length scale smaller than $|a|$. There are, however, additional non-trivial modifications that are not be predicted by
this simple replacement.
Our expressions above explain, for instance, the scaling of the rates with $|r_{\rm eff}|$ used 
in Figs.~\ref{Fig:K3:aN:Bosons} and \ref{Fig:Vrel:Bosons}:
the factor $\sinh 2\eta$ introduces a $|r_{\rm eff}|^{-1}$ suppression.
This reduction of $\eta$, which is connected to transitions to deeply-bound two-body
states, is also responsible for the more pronounced 
minima in Fig.~\ref{Fig:K3:aP:Bosons} as $|r_{\rm eff}|$
increases for those calculations with multiple two-body bound states.
The observation of interference minima in $K_{3}$ is thus more favorable
near a narrow Feshbach resonance, in the sense it will be more pronounced under the same experimental conditions.

Equations~(\ref{Eq:Vrel:aP:Narrow}) and (\ref{Eq:K3:aN:Narrow}) further reveal the fundamental importance of the short-range
three-body physics through their dependence on $A$ in both $\eta$ and $\Phi$.  This physics is absent
from the zero-range treatments~\cite{Petrov2004,Gogolin2008}, so the agreement in Figs.~\ref{Fig:K3:aN:Bosons} 
and~\ref{Fig:Vrel:Bosons} between our numerical
results and the ZRP predictions for the position of the first Efimov feature is rather fortuitous.
We see from the arguments of $\sin^2$ in Eqs.~(\ref{Eq:Vrel:aP:Narrow}) and (\ref{Eq:K3:aN:Narrow}) 
that $A$-independent Efimov feature positions --- as predicted in \cite{Petrov2004,Gogolin2008} --- are found only in the limit
$|{\rm Re} A/r_{\rm eff}|$$\rightarrow$0 and $|{\rm Re} A/r_{\rm eff}|$$\rightarrow$$\infty$.  For the numerical examples
shown above, ${\rm Re} A$$\sim$$r_0$, but this
need not be true in general. In particular, if there is a short-range three-body resonance
near the break-up threshold, the value of ${\rm Re} A$ can, in principle, take any value from $-\infty$ to $+\infty$.

For a broad resonance, the short-range phase $\Phi$ in Eqs.~(\ref{Eq:K3:aP})--(\ref{Eq:Vrel:aP}) corresponds the phase in the hyperradial wavefunction $F_\nu(R)$ at $R$=$r_0$, 
if one assumes that $F_\nu(R)$ takes the following form near $R$=$r_0$:
\begin{equation}
F_\nu(R)\propto \sqrt{R}\sin[s_0\ln(R/r_0)+\Phi+i\eta].
\label{Eq:Radial}
\end{equation} 
For a narrow resonance, the phase $\Phi$ in Eqs.~(\ref{Eq:Vrel:aP:Narrow}) and (\ref{Eq:K3:aN:Narrow}) instead gives the phase of $F_\nu(R)$ at $R$=$\alpha|r_{\rm eff}|$, with $F_\nu(R)$ taking 
the same form as Eq.~(\ref{Eq:Radial}) near $R$=$\alpha|r_{\rm eff}|$. The parameter $\eta$ for a narrow resonance gives an ``effective'' loss parameter as if the loss occurs 
near $R$=$\alpha|r_{\rm eff}|$, with essentially the same physical interpretation as the $\eta$ parameter for a broad resonance. Interestingly, in the limit $\alpha|r_{\rm eff}|$$\rightarrow$$r_0$, 
both $\Phi$ and $\eta$ for a narrow resonance reduce to those for a broad resonance, under the assumption that $\eta\ll 1$. In this limit, we can determine the value of $\beta$ for a broad resonance, 
by comparing the phases in Eqs.~(\ref{Eq:Vrel:aP:Narrow}) and (\ref{Eq:K3:aN:Narrow}) to those in Eqs.~(\ref{Eq:Vrel:aP}) and (\ref{Eq:K3:aN}). This gives $\beta$=$1.4$ for relaxation and 
$\beta$=$2.9$ for recombination.

\subsection{Vibrational relaxation for fermionic system }
\label{Fermions}
To show that the present analysis of narrow resonances is more general than the identical boson system discussed so far, we will consider a two-component Fermi system with a narrow 
interspecies Feshbach resonance.
If the components correspond to the same atomic
isotope in different hyperfine states, say $F$ and $F'$, three-body processes are in general 
suppressed near an interspecies Feshbach resonance.  For recombination, this suppression is a consequence of the Pauli exclusion principle applied to the $FFF'$ system.
In particular, $K_3$ for 
$F+F+F'\rightarrow (FF')^*+F$ is proportional to $k^2$ near zero energy, where $k=\sqrt{2\mu E}$ is the incident wavenumber and $E$ is the collision energy~\cite{Esry2001}. 
Therefore, in the ultracold regime, three-body recombination vanishes for $FFF'$ systems, and we restrict our study to vibrational relaxation since its rate is constant as $k\rightarrow 0$. 
Nevertheless, in this case,  the rate for atom-dimer relaxation, $(FF')^*+F\rightarrow FF'+F$, is suppressed as $1/a^{3.33}$~\cite{DimerPetrov2004,ScalingDIncao2005}, 
an important property since it stables atom-molecule fermionic mixtures.

To get a sense of the effect of large $|r_{\rm eff}|$ on fermionic collisions, we have
studied the atom-dimer relaxation processes near narrow resonances. 
For broad resonances, i.e., $|r_{\rm eff}|\simeq r_{0}$, the usual $a^{-3.33}$ suppression on $V_{\rm rel}^{(F)}$~\cite{DimerPetrov2004} 
originates from a repulsive barrier in the incident adiabatic hyperspherical potential
\begin{eqnarray}
W_{\nu,\nu}(R)\simeq \frac{p_0^2-1/4}{2\mu R^2}
\label{Eq:FermiPot}
\end{eqnarray}
in the range $r_{0}$$\ll$$R$$\ll$$a$~\cite{ScalingDIncao2005}. The universal constant $p_0\approx2.166$ 
is determined by a transcendental equation from the ZRP model~\cite{Efimov1971,Efimov1973}.

\begin{figure}[h]
\begin{center}
\includegraphics[height=1.6in]{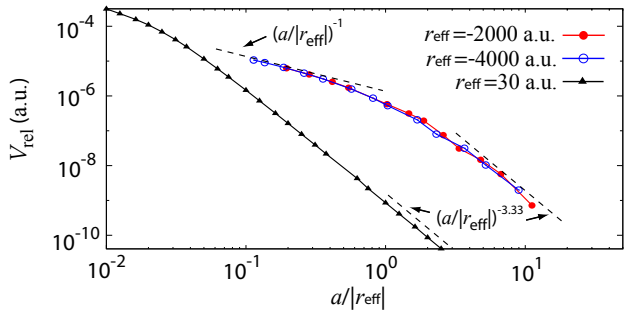}
\end{center}
\caption[Relaxation rates for mixed-spin fermions with $a$$>$0 and large $|r_{\rm eff}|$.]
{\small{Relaxation rates for mixed-spin fermions with $a$$>$0 and large $|r_{\rm eff}|$. The relaxation 
rate for small $|r_{\mathrm{eff}}|$ is also plotted, showing the same scaling with $a$
but not with $|r_{\mathrm{eff}}|$ since it is not in the 
universal limit.
}}
\label{Fig:Fermions}
\end{figure}

When $|r_{\rm eff}|$$\gg$$r_{0}$, however, 
$W_{\nu,\nu}(R)$ is modified in the range $r_{0}$$\ll$$ R$$ \ll $$|r_{\rm eff}|$ 
by the emergence of a Coulomb-like potential, similar to what was found for bosons. However,
the coefficient for the Coulomb-like potential for fermions $c_0\approx 0.3$ and does not change when the number of $s$-wave bound states 
between $F$ and $F'$ is increased.
Nevertheless,  
compared with the potential in Eq.~(\ref{Eq:FermiPot}), the modified repulsive barrier is weakened due to the smallness of the effective charge in the Coulomb-like potential, 
leading to an {\em enhancement} of vibrational relaxation.
Moreover, when $a$$\ll$$|r_{\rm eff}|$, the dependence of the rate on $a$ is
altered, much like for bosons with $a$$\ll$$|r_{\rm eff}|$ in Fig.~\ref{Fig:K3:aP:Bosons}.
All of these effects can be seen in our numerical calculations shown in Fig.~\ref{Fig:Fermions}. 
For $a$$<$$|r_{\rm eff}|$, the numerical result shows that relaxation scales as
$(a/|r_{\rm eff}|)^{-1}$, a much weaker suppression than for broad resonances. 

For $a$$>$$|r_{\rm eff}|$, we 
can apply the same kind of analysis as for bosons, using the fact that the idealized
potential behaves as in Eq.~(\ref{Eq:FermiPot}) for $|r_{\mathrm{eff}}|$$\ll$$ R$$ \ll$$ a$ 
and assuming the potential is zero for $r_0$$\ll$$R$$\ll$$|r_{\rm eff}|$.  
We obtain for this case
\begin{equation}
V_{\rm rel}^{(F)}\!=\!\frac{256 \pi \sqrt{3} p_0^2\,|\mathrm{Im}A|/m}{[(1\!-\!4p_0^2)(\mathrm{Re}A/|\alpha r_{\mathrm{eff}}|)^2\!+
\!(2p_0\!+\!1)^2]^2}\left(\frac{\beta a}{|\alpha r_{\mathrm{eff}}|}\right)^{1\!-\!2p_0}. 
\label{Eq:Vrel:Fermions}
\end{equation}
We recover the broad resonance scaling with $a$, $V_{\rm rel}^{(F)}$$\propto$$(a/|r_{\rm eff}|)^{-3.33}$ when $r_{\rm eff}<0$,
but with a much larger overall magnitude due to the dependence on $r_{\rm eff}$.

\section{Summary}
We have studied ultracold collisions of three identical bosons and of mixed-spin fermions 
near a narrow Feshbach resonance using the connection between $r_{\rm eff}$  for the two-body
interaction and the width of the resonance.
We were able to identify the key modifications to the three-body
adiabatic hyperspherical potentials and thus derive analytical expressions for the
rate constants.  From these analytical expressions, we showed that short-range three-body
physics is still important, even near a narrow Feshbach resonance.  This result is, perhaps,
unfortunate for experimentalists since the positions of the Efimov features are, in general,
still dependent on short-range physics, making it difficult to locate {\em a priori} a 
minimum of $K_3$ as suggested in \cite{Petrov2004}.  On the other hand, our analysis has
shown that bosonic recombination and relaxation to deeply-bound 
two-body states are suppressed near a narrow resonance which might prove beneficial experimentally.  
Similarly, our analysis suggests that long-lived weakly-bound $FF'$ molecules are
most easily obtained near a broad resonance.

\appendix
\section{Adiabatic potentials with the zero-range potential model}
\label{Chap:Narrow:ZRP}
The effective range has been used in Ref.~\cite{Fedorov2001,Jonsell2006} to calculate the regularized three-body adiabatic hyperspherical potentials within ZRP model. 
Here we aim to study the scaling behavior of the potentials with $r_{\rm eff}$
and show the difference between the results from a ZRP treatment and those from numerical calculations near narrow Feshbach resonances.

We extend the ZRP treatment for small effective range~\cite{ZRP} by
including up to the effective-range term in the low-energy expansion of the two-body scattering phase shift
to better represent the scattering properties at finite energy:
\begin{eqnarray}
\frac{\partial}{\partial r_{ij}}(r_{ij} \Psi)=\left (-\frac{1}{a}+\frac{1}{2} r_{\mathrm{eff}} k_{2}^2 \right )\Psi ~~~ ({r_{ij}} \rightarrow 0 )
\label{Eq:ZRP}
\end{eqnarray}
where $r_{ij}$ is the interparticle distance.
Treating $k_{2}^2$ as proportional to two-body kinetic energy operator $\nabla^2_{\boldsymbol{r}_{ij}}$, the above boundary condition leads to the transcendental equation
for the three-body adiabatic equation,
\begin{widetext}
\begin{eqnarray}
s_0\cosh(\frac{\pi}{2}s_0)-\frac{8}{\sqrt{3}}\sinh(\frac{\pi}{6}s_0)
=12^{-\frac{1}{4}}\sinh(\frac{\pi}{2}s_0)\left( \frac{2R}{a}+\frac{r_{\mathrm{eff}}}{R}s_0^2\right),
\end{eqnarray}
\end{widetext}
where $s_0$ is related to the adiabatic hyperspherical potential by 
\begin{eqnarray}
U_\nu=-\frac{s_0^2+\frac{1}{4}}{2\mu R^2}. 
\end{eqnarray}
As mentioned earlier, in the ZRP model, the diagonal coupling $Q_{\nu,\nu}/2\mu$
is of higher order than $1/R^2$ for $R\ll |a|$ and is thus negligible in this region.  
The lowest effective adiabatic potential $W_{0,0}(R)$ then behaves like 
\begin{eqnarray}
W_{0,0}(R)\simeq -\frac{c_0}{2\mu |r_{\mathrm{eff}}| R}-\frac{1/4}{2\mu R^2},
\label{Eq:Pot:ZRP:Narrow}
\end{eqnarray}
in the region $r_0\ll R\ll |r_{\mathrm{eff}}|$ with $c_0\approx 1.68$.
Due to the largeness of $|r_{\mathrm{eff}}|$, the potential in Eq.~(\ref{Eq:Pot:ZRP:Narrow}) is dominated by the $1/R^2$ term.
The form of the zero-range three-body potential thus disagrees with the numerical result for the Coulomb-like potential by having an additional $1/R^2$.
Further, the non-universal property of the three-body potentials seen in the numerical calculations in this region is not manifested in the ZRP potential at all.
The behavior of the potentials for the continuum channels in the region $r_0\ll R\ll |r_{\mathrm{eff}}|$, however, agrees with the numerical results. 
They both behave like the potentials for free particles:
\begin{eqnarray}
W_{\nu,\nu}=\frac{\lambda(\lambda+4)+15/4}{2\mu R^2},
\end{eqnarray}
where $\lambda$ is a non-negative integer~\cite{Esry2001}.  
In the Efimov region $|r_{\mathrm{eff}}|\ll R\ll |a|$ and asymptotic
region $R\gg |a|$, the leading behavior of the zero-range three-body potentials agrees with the numerical results, as they both reduce to 
the behavior for small $|r_{\mathrm{eff}}|$ case.

\section{Table of effective range for selected resonances}
\label{Append:Table}
We list the effective range and short-range radius ($r_0$) for some narrow Feshbach resonances in Table~\ref{Tab:EffRange}. 
The effective ranges are calculated using the resonance parameters from Ref.~\cite{FeshbachChin}. The short-range radius $r_0$ is given 
by the van der Waals length~\cite{Braaten2006}.
From Fig.~\ref{Fig:K3:aP:Bosons} and \ref{Fig:Vrel:Bosons} we observe that
the ZRP results for scattering observables give reasonably predict the first Efimov features only when $|r_{\rm eff}|$ is beyond a few thousand.
For the resonances listed in Table~\ref{Tab:EffRange}, this implies that the ZRP results may only be applied for those with width below 1~G, given Re$A\approx r_0$ and $r_0=50$~a.u..
\begin{center}
  \begin{table*}
    \begin{tabular}[c]{lllllll}
      \hline\hline
      Atomic species & Res. position (G) & $a_{\rm bg}$ (a.u.) & $\Delta\mu$ ($\mu_B$) & $\Delta B$ (G) & $r_{\rm eff}$ (a.u.) & $r_0$\\
      \hline
        $^6$Li & 543.25  & 60 & 2 & 0.1 & --71300 & 62.5\\
        $^{23}$Na & 1195  & 62 & --0.15 & 1.4 & --17100 & 89.9\\
        $^{23}$Na & 907  & 63 & 3.8 & 1 & --947 & ''\\
        $^{23}$Na & 853  & 63 & 3.8 & 0.0025 & --373000 & ''\\
        $^{87}$Rb & 1007.2 & 100 & 2.79 & 0.21 & --1010 & 165\\
        $^{87}$Rb &  911.7 & 100 & 2.71 & 0.0013 & --168000 & ''\\
        $^{87}$Rb &  685.4 & 100 & 1.34 & 0.006 & --73400 & ''\\
        $^{87}$Rb &  406.2 & 100 & 2.01 & 0.0004 & --734000 & ''\\
        $^{87}$Rb &  9.13 & 99.8 & 2.00 & 0.015 & --19700 & ''\\
        $^{133}$Cs &  47.97 & 926 & 1.21 & 0.12 & --287 & 202\\
        $^{133}$Cs &  19.84 & 160 & 0.57 & 0.005 & --84600 & ''\\
        $^{133}$Cs &  53.5 & 995 & 1.52 & 0.0025 & --10200 & ''\\
        $^{52}$Cr &  589.1 & 105 & 2 & 1.7 & --276 & 91.3\\
        $^{52}$Cr & 499.9 & 107 & 4 & 0.08 & --2880 & ''\\
        $^{39}$K+$^{87}$Rb & 317.9 & 34 & 2 & 7.6 & --185 & 143\\
      \hline
      \hline
    \end{tabular}
    \caption[Effective range for some selected Feshbach resonances.]
    {{Effective-range for some selected Feshbach resonances.}}
    \label{Tab:EffRange}
  \end{table*}
\end{center}

\section{Single-channel approach for deriving three-body inelastic rates}
\label{Append:Analytic}

To derive Eqs.~(\ref{Eq:Vrel:aP:Narrow}), (\ref{Eq:K3:aN:Narrow}) and (\ref{Eq:Vrel:Fermions}), we use a variation of the optical potential approach~\cite{Friedrich}. Optical potentials, 
which are non-Hermitian, have long been added to otherwise Hermitian Hamiltonians to allow flux to be lost to degrees of freedom not explicitly treated. The original 
Feshbach projection formalism~\cite{Friedrich} dictates the rigorous way to do this, but often the optical potential is introduced phenomenologically.

Three-body recombination and vibrational relaxation necessarily involve multiple adiabatic hyperspherical potentials and can be calculated within a multichannel
scattering formalism. For recombination with $a<0$ and relaxation with $a>0$, however, all of the universal behavior is determined by the initial channel alone 
since the non-adiabatic couplings --- and thus the inelastic transitions --- occur predominantly in the non-universal region $R\lesssim r_0$. We can thus use 
the optical potential approach to replace all of the deeper-lying final adiabatic channels in these cases by short-range phenomenological parameters. Instead of actually 
modifying $W_{\nu,\nu}$, though, we will use the equivalent approach of imposing a complex boundary condition at $R=r_0$~\cite{Jonsell2006} where the imaginary part is related 
to the inelastic transition probability.

We thus begin with the schematic potential described in Fig.~\ref{Fig:NarrowPots:Ideal}:
\begin{align}
W_{\nu,\nu}(R)=\left\{
\begin{array}{cc}
\displaystyle 0 &  r_0<R<\alpha|r_{\rm eff}|,\\
\displaystyle -\frac{s_0^2+1/4}{2\mu R^2} &  \alpha|r_{\rm eff}|<R<\beta|a|,\\
\displaystyle E_{\nu}+\frac{l(l+1)}{2\mu R^2} & R>\beta|a|
\end{array}\right.,
\label{Eq:NarrowPotential}
\end{align}
where the threshold energy for the initial channel $E_{\nu}$ is zero for recombination and $-1/2\mu_2a^2$ for relaxation.
The asymptotic , free-particle potential is characterized by $l$=$0$ for relaxation and $l$=$3/2$ for recombination. The 
parameters $\alpha$ and $\beta$, expected to be universal and in the order of unity, represent the fact that the boundaries 
between regions are not so sharply defined in practice. They can be fixed by fitting the final analytic expressions to numerical rates.
By neglecting the small residual non-adiabatic couplings between the initial and the final channel for $R>r_0$, the hyperradial wavefunction in the initial channel 
can be written down piece-wise:
\begin{widetext}
\begin{align}
F_\nu(R)=\left\{
\begin{array}{lc}
C_1\sin[k R+\delta_s(k)] & r_0< R< \alpha|r_{\rm eff}|,\\
C_2\sqrt{R}[J_{i s_0}(kR)+\tan\delta_2 N_{i s_0}(kR)] & \alpha|r_{\rm eff}|< R< \beta|a|,\\
C_3\sqrt{R}[J_{l+1/2}(kR)+\tan\delta N_{l+1/2}(kR)] & R> \beta|a|, 
\label{Eq:AsymSoln}
\end{array}\right.
\end{align}
\end{widetext}
For $k\rightarrow 0$ and $|a|\gg |r_{\rm eff}|$, the short-range phase shift $\delta_s$ is 
\begin{eqnarray}
\delta_s(k)=-A k,
\end{eqnarray}
where $A$ is the short-range three-body scattering length introduced in Sec.~\ref{Potentials}. Due to the inelastic transitions at $R\leq r_0$, $A$ acquires an imaginary part which
determines the strength of the transition.
Matching the hyperradial wavefunction $F_0(R)$ at $\alpha|r_{\rm eff}|$ and $\beta|a|$ gives the asymptotic phase shift $\delta$ 
in terms of $\alpha|r_{\rm eff}|$, $\beta|a|$ and $A$. The probability for elastic scattering ${\cal R}$ is the reflection coefficient:
\begin{eqnarray}
{\cal R}=\left|\frac{1+i\tan\delta}{1-i\tan\delta} \right|^2.
\end{eqnarray}
The probability for an inelastic transition is then determined as
\begin{widetext}
\begin{eqnarray}
1-{\cal R}=\frac{2\pi}{\Gamma(l+\frac{3}{2})\Gamma(l+\frac{1}{2})}\left(\frac{k\beta|a|}{2} \right)^{2l+1}\frac{\sin2\varphi_0\sinh2\eta}{\sinh^2\eta+\sin^2[s_0\ln(|a/r_{\mathrm{eff}}|)+\Phi+\varphi]},
\end{eqnarray}
\end{widetext}
where the parameters $\Phi$,  $\eta$, $\varphi$ and $\varphi_0$ are defined in Eqs.~(\ref{Eq:Narrow:Phi})--(\ref{Eq:Narrow:Varphi}).
The recombination rate $K_3$ and the relaxation rate $V_{\rm rel}$ are expressed by the inelastic transition probability through
\begin{align}
K_3&=\frac{192\pi^2}{\mu k^2}{(1-{\cal R})},\\
V_{\rm rel}&=\frac{\pi}{\mu k} (1-{\cal R}).
\end{align}

For the mixed-spin fermionic system $FFF'$, we are interested in the three-body relaxation process $(FF')^*+F\rightarrow FF'+F$, where $(FF')^*$ is a weakly-bound
molecule and $FF'$ is a deeply-bound molecule. Since the adiabatic hyperspherical potential for this system is different from the potential for bosons only in the region
$\alpha|r_{\rm eff}|< R< \beta|a|$, we can calculate $FFF'$ recombination with this same formalism.
Thus, $W_{\nu,\nu}$ in $\alpha|r_{\rm eff}|<R<\beta |a|$ must be replaced by 
\begin{eqnarray}
W_{\nu,\nu}(R)= \frac{p_0^2-1/4}{2\mu R^2},
\end{eqnarray}
where $p_0$$$$$ is the universal constant given in Sec.~\ref{Fermions}.

The corresponding hyperradial wavefunction is thus changed to 
\begin{eqnarray}
F_\nu(R)=C_2\sqrt{R}[J_{p_0}(kR)+\tan\delta_2 N_{p_0}(kR)],
\end{eqnarray}
for $\alpha |r_{\rm eff}|< R <\beta|a|$.
Following exactly the same analysis as for identical bosons, we get the relaxation rate for $FFF'$ system Eq.~(\ref{Eq:Vrel:Fermions}).

\begin{acknowledgements}
This work was supported by the National Science Foundation. 
\end{acknowledgements}

\bibliography{References}{}
\bibliographystyle{apsrev}

\end{document}